\documentclass[a4paper,10pt]{article}
\usepackage{amssymb,amsfonts,amsmath,mathtools}
\usepackage{rotating,array,ragged2e}
\usepackage{breakurl}
\usepackage{xr}
\usepackage[all]{xy}
\usepackage[authoryear]{natbib}
\usepackage[breaklinks=true]{hyperref}

\newcounter{statements}
\newcounter{examples}

\newcommand*\oldurlbreaks{}
\let\oldurlbreaks=\UrlBreaks
\renewcommand{\UrlBreaks}{\oldurlbreaks\do\a\do\b\do\c\do\d\do\e%
  \do\f\do\g\do\h\do\i\do\j\do\k\do\l\do\m\do\n\do\o\do\p\do\q%
  \do\r\do\s\do\t\do\u\do\v\do\w\do\x\do\y\do\z
  \do\A\do\B\do\C\do\D\do\E\do\F\do\G\do\H\do\I\do\J\do\K\do\L
  \do\M\do\N\do\O\do\P\do\Q\do\R\do\S\do\T\do\U\do\V\do\W\do\X
  \do\Y\do\Z\do\?\do\&\do\_\do\%}

\renewcommand{\ln}{\textnormal{ln}}

\renewcommand{\ker}{\textnormal{ker}}

\newcommand{\spn}{\textnormal{span}}

\newcommand{\defbf}[1]{\textnormal{\textbf{#1}}}

\newcommand{\myspace}{\text{\ }}

\newcommand{\aset}{\mathcal{A}}

\newcommand{\sset}{\mathcal{S}}

\newcommand{\cset}{\mathcal{C}}

\newcommand{\rset}{\mathcal{R}}

\newcommand{\kset}{\mathcal{K}}

\newcommand{\pset}{\mathcal{P}}

\newcommand{\dset}{\mathcal{D}}

\newcommand{\uset}{\mathcal{U}}

\newcommand{\vset}{\mathcal{V}}

\newcommand{\eset}{\mathcal{E}}

\newcommand{\yset}{\mathcal{Y}}

\newcommand{\crn}{(\sset, \cset, \rset, \kset)}

\newcommand{\rn}{(\sset, \cset, \rset)}

\newcommand{\spacereactioncolumn}{\text{\ \ \ }}

\newcommand{\myrelc}{\text{\raisebox{-0.04cm}{$\stackrel{\scriptscriptstyle\mathcal{C}}{\leftrightsquigarrow}$}}}

\newcommand{\myrelcsmall}{\text{\raisebox{-0.08cm}{\scalebox{1}{$\stackrel{\scriptscriptstyle\mathcal{C}}{\leftrightsquigarrow}$}}}}

\newcommand{\myrelcsmallx}[1]{\text{\raisebox{-0.08cm}{\scalebox{1}{$\stackrel{\scriptscriptstyle\mathcal{C}}{\leftrightsquigarrow}^{_{#1}}$}}}}

\newcommand{\myrelrsmall}{\text{\raisebox{-0.08cm}{\scalebox{1}{$\stackrel{\scriptscriptstyle\mathcal{R}}{\leftrightsquigarrow}$}}}}

\newcommand{\myrelrsmallx}[1]{\text{\raisebox{-0.08cm}{\scalebox{1}{$\stackrel{\scriptscriptstyle\mathcal{R}}{\leftrightsquigarrow}^{_{#1}}$}}}}

\newcommand{\rii}{\tilde{y}\rightarrow\tilde{y}'}
\newcommand{\ri}{y\rightarrow y'}

\newcommand{\yi}{y}
\newcommand{\yii}{\tilde{y}}

\newcommand{\viiaster}{[v(c^*)]_{\rii}}
\newcommand{\viaster}{[v(c^*)]_{\ri}}
\newcommand{\vii}{[v(c)]_{\rii}}
\newcommand{\vi}{[v(c)]_{\ri}}

\newcommand{\mypartr}{\rset/\myrelrsmall}
\newcommand{\mypartc}{\cset/\myrelcsmall}

\newcommand{\myvert}{\myspace\vert\myspace}
\newcommand{\myand}{\myspace\text{and}\myspace}

\newcommand{\mybig}[1]{\raisebox{-0.04cm}{\resizebox{.025\hsize}{!}{$#1$}}}

\newcommand{\mybigg}[1]{\raisebox{-0.08cm}{\resizebox{.035\hsize}{!}{$#1$}}}

\newcommand{\equationcolsep}{\renewcommand{\arraycolsep}{0.05cm}}

\newcommand{\oderowsep}{\renewcommand{\arraystretch}{1.3}}

\newcommand{\myconc}[2]{
	#1_{\text{#2}}
}

\newcommand{\mydifftext}[1]{
	\text{d}#1/\text{d}t
}

\newcommand{\myzero}{
	\mathbf{0}
}

\newcommand{\bw}{
	0.5cm
}

\newcommand{\bx}{
	0.7cm
}

\newcommand{\w}{
	0.46cm
}

\newcommand{\wsmall}{
	0.25cm
}

\newcolumntype{x}[1]{
>{\raggedleft\hspace{0pt}$}p{#1}<{$}}

\newcolumntype{C}[1]{>{\centering\arraybackslash}m{#1}}

\def\ra#1{\rotatebox{90}{\begin{minipage}[t][0.2cm][t]{0.5cm}$#1$\end{minipage}}}

\newcommand{\statement}[5]
{
	\bigskip
	\refstepcounter{#3}\label{#2}\noindent\textbf{#1 \ref{#2}.} \textit{#4}

	\proof{#5}
}

\newcommand{\proof}[1]
{
	\text{\ }\\
	\noindent\textbf{Proof.}
	#1\qed
}

\newcommand{\lemma}[4]
{
	\statement{Lemma}{#1}{#2}{#3}{#4}
	\bigskip
}

\newcommand{\definition}[3]
{
	\bigskip
	\refstepcounter{#2}\label{#1}\noindent\textbf{Definition \ref{#1}.} \textit{#3}
	\bigskip
}

\newcommand{\example}[3]
{
	\bigskip
	\bigskip
	\refstepcounter{#2}\label{#1}\fbox{
		\parbox{0.9\linewidth}{
			\medskip\noindent\textbf{Example \ref{#1}.} #3\medskip
		}
	}
	\bigskip
	\bigskip
}

\newcommand{\lemmasupplement}[4]
{
	\statementsupplement{Lemma}{#1}{#2}{#3}{#4}
	\bigskip
}

\newcommand{\corollarysupplement}[4]
{
	\statementsupplement{Corollary}{#1}{#2}{#3}{#4}
	\bigskip
}

\newcommand{\definitionsupplement}[3]
{
	\bigskip
	\refstepcounter{#2}\label{#1}\noindent\textbf{Definition S\ref{#1}.} \textit{#3}
	\bigskip
}

\newcommand{\statementsupplement}[5]
{
	\bigskip
	\refstepcounter{#3}\label{#2}\noindent\textbf{#1 S\ref{#2}.} \textit{#4}

	\proof{#5}
}

\newcommand{\myarraytwo}[5]
{
	\begin{array}[h]{crc}
		&
		\left.
		\equationcolsep
		\begin{array}{#2}
			#4
		\end{array}
		\right.\text{\ } & \\
		& & \\
		#1 &
		\left[
		\equationcolsep
		\begin{array}[h]{#2}
			#3
		\end{array}
		\right] &
		\equationcolsep
		\begin{array}{l}
			#5
		\end{array}
	\end{array}
}

\newcommand{\myarraythree}[4]
{
	\begin{array}[h]{crc}
		#1 &
		\left[
		\equationcolsep
		\begin{array}[h]{#2}
			#3
		\end{array}
		\right] &
		\equationcolsep
		\begin{array}{l}
			#4
		\end{array}
	\end{array}
}

\def\qed {{
   \parfillskip=0pt        
   \widowpenalty=10000     
   \displaywidowpenalty=10000  
   \finalhyphendemerits=0  
   \leavevmode             
   \unskip                 
   \nobreak                
   \hfil                   
   \penalty50              
   \hskip.2em              
   \null                   
   \hfill                  
   $\square$
   \par}}                  

\begin{document}

\begin{titlepage}
\author{\\Jost Neigenfind$^{1,\ast}$, Sergio Grimbs$^{2}$ and Zoran Nikoloski$^{1}$\\
\text{\ }\\
$^{1}$Max-Planck Institute of Molecular Plant Physiology\\
$^{2}$Institute of Biochemistry and Biology, University of Potsdam\\
Potsdam, Germany\\
\text{\ }\\
$^*$To whom correspondence should be addressed.\\
E-mail: Neigenfind@mpimp-golm.mpg.de} 
\title{\Large \textbf{On the relation between reactions and complexes of (bio)chemical reaction networks}}
\date{} 
\maketitle
\thispagestyle{empty}
\end{titlepage}

\setcounter{page}{1}

\newpage
\begin{abstract}
Robustness of biochemical systems has become one of the central questions in systems biology although it is notoriously difficult to formally capture its multifaceted nature.
Maintenance of normal system function depends not only on the stoichiometry of the underlying interrelated components, but also on a multitude of kinetic parameters.
Invariant flux ratios, obtained within flux coupling analysis, as well as invariant complex ratios, derived within chemical reaction network theory, can characterize robust properties of a system at steady state.
However, the existing formalisms for the description of these invariants do not provide full characterization as they either only focus on the flux-centric or the concentration-centric view.
Here we develop a novel mathematical framework which combines both views and thereby overcomes the limitations of the classical methodologies.
Our unified framework will be helpful in analyzing biologically important system properties. 
\end{abstract}

\newpage
\tableofcontents

\newpage
\section{Introduction}
	Biochemical networks have evolved to operate in the face of internal and external perturbations \citep{Kitano2004a}.
	The response to these perturbations has shaped the systemic architectural blueprint comprising multiple layered and interrelated components (e.g., genes, proteins, metabolites).
	The dynamic processes involving network-related biochemical components depend on a multitude of kinetic parameters, which remain elusive even for medium-size systems.
	Therefore, methods establishing a connection between structure and dynamics of biochemical systems hold the promise to enable the rigorous study of processes taking place on the underlying biochemical networks both at steady-state as well as dynamic setting.
	
	Two different classes of approaches have been developed to facilitate para-meter-independent analysis of biochemical networks:
	(\textit{i}) flux-focused approaches, including: flux balance analysis (FBA) \citep{Varma1994} and its derivatives \textendash flux variability analysis (FVA) \citep{Mahadevan2003} and flux coupling analysis (FCA) \citep{Burgard2004, Marashi2010}, elementary flux modes (EFMs) \citep{Schuster2000}, and extreme pathways \citep{Schilling1999};
	and (\textit{ii}) concentration-centric approaches, rooted in chemical reaction network theory (CRNT) \citep{Horn1972, Feinberg1979, Feinberg1995} and stoichiometric network analysis \citep{Clarke1988}.

	Given a biochemical network, FBA relies on a linear programming formulation to calculate the steady-state fluxes under the assumption that the investigated organism operates towards optimizing an objective function (e.g., optimizing yield for metabolic networks \citep{Varma1994}).
	FVA also has a linear programming formulation, with the aim of calculating the minimum and maximum values of individual steady-state fluxes for a particular value of the objective.
	FCA can be used to determine pairs of reactions whose flux ratio is the same in each steady state under the same environmental conditions.
	Like FBA and FVA, this approach can also be cast as a linear program.
	On the other hand, approaches based on EFMs allow decomposition of a given network into its smallest functional units operating in a steady state \citep{Schuster2000, Schilling1999}.
	Although the problem of determining the set of all EFMs for a given biochemical network is computationally demanding, recent parallelized implementations of algorithms for EFM computation facilitate EFM-based analysis of genome-scale metabolic networks \citep{Terzer2008}.
	Essential to both flux-based approaches is the usage of the underlying stoichiometric matrix which, without a specified kinetics, cannot be employed to make statements about steady-state metabolite concentrations.

	In contrast, CRNT uses mass-action formulation to study the qualitative behavior of the steady-state concentrations of the components regardless of the parameter values, i.e., for all steady-state reaction fluxes of the mass-action system satisfying the constraints imposed by the stoichiometry. 
	The results of this framework answer questions related to the possibility for existence of multiple steady states, and rely on a structural index determined by interleaving the graph-theoretic and stoichiometric descriptions of the investigated network \citep{Horn1972, Feinberg1979, Feinberg1995, Gunawardena2003, Conradi2007}.

	Biochemical network invariants are of particular interest specifically because they relate to the principle of homeostasis.
	For instance, under the steady-state assumption, the concentrations of components do not change and, thus, are invariant.
	However, invariants in biochemical networks can be defined not only with respect to changes over time, but also changes with respect to different steady states that the system may assume under same environmental conditions (i.e., initial conditions and/or constraints).
	Note that the latter excludes the analysis of trivial invariants which are imposed in the form of conservation relations \citep{Schilling1999, Schuster1996}.
	
	In other words, invoking the steady-state assumption may induce additional invariants with respect to individual components or their combinations, which can ultimately reveal possible reduction in complexity of the system.
	As already stated, FCA provides the means for determining pairs of reactions whose ratio of fluxes is the same in each steady state the system may assume.
	
	In general, changes in fluxes and concentration, as key descriptors of the transitional behavior in biochemical networks, depend on each other.
	This stems from the fact that the reaction rate, i.e., flux, is cast as a function of the concentrations of the considered components.
	Therefore, the question arises whether there exist invariants on the level of concentrations and, if so, whether there is a connection between flux- and concentration-invariants.
	The answer to this question of course depends on the choice of kinetic law providing the relation between reaction fluxes and concentrations.

	Here, we focus on mass action kinetics, representing the simplest and most fundamental law of kinetics, to establish a connection between flux and con-centration-invariants.
	By interleaving the flux- and concentration-invariants, we provide a fundamentally new theoretical approach which can be used to uncover dependencies between fluxes and between concentrations, ultimately leading to a better understanding of system complexity.
	
	Therefore, our study establishes a connection between the two different views of computational systems biology \textemdash the flux-centric and the concentration-centric view.
	Since the theories and methods pertaining to the two views use different notations, a brief overview is provided to describe the used notation. 
	
\section{Methods}
	In chemistry, the law of mass action was established by Guldberg and Waage in the nineteenth century \citep{Guldberg1899, Abrash1986a}.
	It assumes a mixture of large numbers of components which are homogeneously distributed, allowing approximation of the components' behavior with continuous variables.
	A reversible reaction, i.e., a reaction which can proceed in the forward and backward direction, is split into two reactions \textemdash the irreversible forward reaction and the irreversible backward reaction.
	The components consumed by an irreversible reaction are called \textit{substrates}, while those produced are referred to as \textit{products}.
	A reaction's rate is then modeled to be proportional to the product of the concentrations of the participating substrates, especially in the case of an elementary reaction which cannot be further divided into intermediate steps \citep[pg. 385]{Moore1986}.
	Under realistic chemical conditions, it is often the case that a given reaction almost certainly proceeds in one direction.
	In this situation, with the assumption that the reaction rate in one of the directions can be neglected, the reaction is treated as irreversible.
	Therefore, most models of biochemical networks consist of a mixture of reversible and irreversible reactions.

	Here, for the application of specific theoretical methodology, each biochemical network must be transformed to an equivalent one that consists only of irreversible reactions.
	Such a transformation is performed as follows \citep{Gagneur2004}:
 	Let the complete set of reactions be denoted by $\mathcal{R} = \mathcal{R}_{irr} \cup \mathcal{R}_{rev}$, where $\mathcal{R}_{irr}$ denotes the subset of irreversible reactions and $\mathcal{R}_{rev}$ the subset of reversible reactions.
	The set of reactions $\mathcal{R}'_{irr}$ is derived by splitting each reversible reaction from $\mathcal{R}_{rev}$ into two irreversible reactions, one in each direction.
	The original network can then be described by a new set of reactions $\mathcal{R}' = \mathcal{R}'_{irr} \cup \mathcal{R}_{irr}$ with $|\mathcal{R}'| = 2 |\mathcal{R}_{rev}| + |\mathcal{R}_{irr}|$.
	The starting point for our methodologies derived here is always a biochemical network which is of this form, i.e., we assume that $\mathcal{R}$ denotes a set of irreversible reactions (see Example \ref{example_reactions}).

	\example{example_reactions}{examples}{
		The eight irreversible reactions in the set $\mathcal{R} = \{R_1,R_2,R_3,R_4,R_5,R_6,R_7,R_8\}$, given by
		\begin{equation*}
			\label{chapter:introduction:equation:homer:network:general}
			\equationcolsep
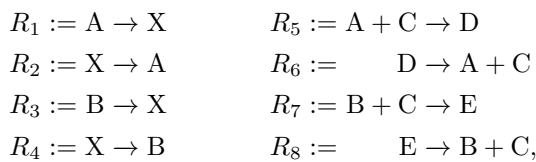
			\begin{array}[h]{ccc}
				\equationcolsep
				\oderowsep
				\begin{array}[h]{rrcl}
					R_1 := & \text{A}            & \xrightarrow[]{} & \text{X}\\
					R_2 := & \text{X}            & \xrightarrow[]{} & \text{A}\\
					R_3 := & \text{B}            & \xrightarrow[]{} & \text{X}\\
					R_4 := & \text{X}            & \xrightarrow[]{} & \text{B}
				\end{array} & \spacereactioncolumn\spacereactioncolumn\spacereactioncolumn &
				\equationcolsep
				\oderowsep
				\begin{array}[h]{rrcl}
					R_5 := & \text{A} + \text{C} & \xrightarrow[]{} & \text{D}\\
					R_6 := & \text{D}            & \xrightarrow[]{} & \text{A} + \text{C}\\
					R_7 := & \text{B} + \text{C} & \xrightarrow[]{} & \text{E}\\
					R_8 := & \text{E}            & \xrightarrow[]{} & \text{B} + \text{C},
				\end{array}
			\end{array}
		\end{equation*}
		can in fact be regarded as four reversible reactions.
		The reversible reactions are formed by $R_1$ and $R_2$, $R_3$ and $R_4$, $R_5$ and $R_6$ as well as $R_7$ and $R_8$.
	}
	
 	The results from flux-centric approaches rely on investigating vector spaces associated to the stoichiometric matrix $N$ (see Example \ref{example_stoichiometric_matrix}).
 	The principal object in the flux-centric approaches is given by the reactions and their fluxes.
 	Here the term ``flux'' is used synonymously to ``reaction rate''.
	A crucial vector space is that of the kernel of the stoichiometric matrix $N$, which is represented by the set of flux vectors $v$ fulfilling $Nv = \myzero$.
	Thus, the kernel of $N$ describes all possible steady-state fluxes of the considered biochemical system.

	\example{example_stoichiometric_matrix}{examples}{
		The set of reactions from Example \ref{example_reactions} give rise to the following stoichiometric matrix:
		\begin{equation*}
			\myarraytwo{N =}{x{\bx}x{\bx}x{\bx}x{\bx}x{\bx}x{\bx}x{\bx}x{\bx}}{
				-1 &  1 &  0 &  0 & -1 &  1 &  0 &  0\tabularnewline
				 0 &  0 & -1 &  1 &  0 &  0 & -1 &  1\tabularnewline
				 0 &  0 &  0 &  0 & -1 &  1 & -1 &  1\tabularnewline
				 0 &  0 &  0 &  0 &  1 & -1 &  0 &  0\tabularnewline
				 0 &  0 &  0 &  0 &  0 &  0 &  1 & -1\tabularnewline
				 1 & -1 &  1 & -1 &  0 &  0 &  0 &  0
			}{
			
				\ra{R_1} & \ra{R_2} & \ra{R_3} & \ra{R_4} & \ra{R_5} & \ra{R_6} & \ra{R_7} & \ra{R_8}
			}{
				\text{A}\\
				\text{B}\\
				\text{C}\\
				\text{D}\\
				\text{E}\\
				\text{X}
			}.
		\end{equation*}
	}

	The concentration-centric approaches, represented by CRNT, use a notation which combines linear algebra and set theory \citep{Gunawardena2003}.
	For a given set of reactions, the set of complexes $\mathcal{C}$ is composed of the left- and right-hand sides of each reaction arrow.
	Any reaction $y\rightarrow y'\in\mathcal{R}$ can then easily be defined in terms of its complexes $y,y'\in\mathcal{C}$.

	Results from CRNT establish a relationship between the structure of a mass action system's network and the nature of the set of equilibria of the corresponding system of ODEs, independently of the rate constants \citep{Feinberg1995}.
	Let $\mathbb{P} = \{x\in \mathbb{R}\myspace|\myspace x > 0\}$ be the set of	positive real numbers.
	In the following, it is assumed that, if the system of ODEs of a mass action system admits an equilibrium, then the species' concentrations satisfy the following condition: 

	\definition{chapter_paper_def_posstst}{statements}{
		Let $c$ be the vector of concentrations of a mass action system.
		The system admits a \defbf{positive steady state} if $\mydifftext{c} = \myzero$ and $c \in\mathbb{P}$.
	}

	A reaction network which mathematically captures the graph-theoretic properties of a chemical mass action system is defined as follows \citep{Feinberg1979, Feinberg1995, Gunawardena2003}: 

	\definition{reaction_network}{statements}{
	   	A \defbf{reaction network} is a triple $\rn$ where $\sset$ is a finite set of species;
		$\cset$ is a finite set of multisets of species, called complexes;
		$\rset$ is a relation on $\cset$, denoted by $\ri$ for $y,y' \in \cset$, which represents a reaction converting $y$ to $y'$.
	}

	Based on a reaction network, a chemical reaction can be defined as follows \citep{Feinberg1979, Feinberg1995, Gunawardena2003} (see also Example \ref{example_chemical_reaction_network}): 
	
	\definition{chemical_reaction_network}{statements}{
	   	A \defbf{chemical reaction network} $\crn$ is a reaction network endowed with a function $\kset: \rset\rightarrow \mathbb{P}$ which associates a positive rate constant to each reaction of the reaction network.
	}

	\example{example_chemical_reaction_network}{examples}{
		The set of reactions in Example \ref{example_reactions} gives rise to the chemical reaction network $(\mathcal{S},\mathcal{C},\mathcal{R},\mathcal{K})$ with
		\begin{equation*}
			\mathcal{S} = \{\text{A},\text{B},\text{C},\text{D},\text{X}\},
		\end{equation*}
		representing the set of species,
		\begin{equation*}
			\mathcal{C}=\mybig{\{}\{\text{A}\},\{\text{B}\},\{\text{A},\text{C}\},\{\text{B},\text{D}\},\{\text{D}\},\{\text{E}\},\{\text{X}\}\mybig{\}},
		\end{equation*}
		representing the set of complexes,
		\begin{equation*}
			\mathcal{R} = \{R_1,R_2,R_3,R_4,R_5,R_6,R_7,R_8\},
		\end{equation*}
 		representing the set of reactions, and
		\begin{equation*}
			\mathcal{K} = \{k_{R_1},k_{R_2},k_{R_3},k_{R_4},k_{R_5},k_{R_6},k_{R_7},k_{R_8}\},
		\end{equation*}
		representing the set of rate constants.
	}

	In the following, we use the concept of a partition of a set (see Definitions S\ref{supp_definition_equivalence_relation} - S\ref{supp_definition_partition} in the Appendix) to obtain new insights into the dynamics of chemical reaction networks.
	Of particular interest are the partitions of the set of reactions and those of the set of complexes (see also Example \ref{example_general_partitions}), both of which can be placed in a relation defined as follows:

	\definition{definition_partition_less_or_equal}{statements}{
		Given a set $\aset$ and two equivalence relations $\thicksim$ and $\thicksim'$ on the elements of $\aset$ such that $\forall\dset'\in\aset/\thicksim'\exists\dset\in\aset/\thicksim$ with $\dset'\subseteq\dset$.
		Then, $\aset/\thicksim$ is said to be a \defbf{coarser} partition than $\aset/\thicksim'$ and $\aset/\thicksim'$ is said to be a \defbf{finer} partition than $\aset/\thicksim$, denoted by $\aset/\thicksim'\hspace{0.1cm}\leq\hspace{0.1cm} \aset/\thicksim$.
	}

	\example{example_general_partitions}{examples}{
		\begin{itemize}
		\item[(a)] The set $\rset/\thicksim  = \{\{R_1,R_2,R_3,R_4\},\{R_5,R_6,R_7,R_8\}\}$ satisfies the conditions of Definition S\ref{supp_definition_partition}.
				   Therefore, $\{R_1,R_2,R_3,R_4\}$ and $\{R_5,R_6,R_7,R_8\}$ are equivalence classes since $R_1\thicksim R_2\thicksim R_3\thicksim R_4$ and $R_5\thicksim R_6\thicksim R_7\thicksim R_8$.
		\item[(b)] The set $\rset/\thicksim' = \{\{R_1,R_2\},\{R_3,R_4\},\{R_5,R_6,R_7,R_8\}\}$ also satisfies the conditions of Definition S\ref{supp_definition_partition}.
		\item[(c)] The set $\{\{R_1,R_2,R_3,R_4,R_5\},\{R_5,R_6,R_7,R_8\}\}$ does not satisfy condition (ii) of Definition S\ref{supp_definition_partition} and, therefore, does not represent a partition of $\rset$.
		\item[(d)] By Definition \ref{definition_partition_less_or_equal} it is not difficult to see that $\rset/\thicksim'\hspace{0.1cm}\leq\hspace{0.1cm} \rset/\thicksim$.
		\end{itemize}		
	}

\section{Results}
	\subsection{Invariant reaction ratios}
		Consider two reactions $\ri, \rii\in\rset$ whose rate ratio at each positive steady state $c$ is an invariant, i.e., $\vi/\vii = $ const.
		Such pairs of reactions can readily be obtained by solving a linear program typical to FCA.
		We note that in FCA, a pair of reactions satisfying $\vi/\vii = $ const is referred to as \textit{fully coupled} reactions.
		Then, for two positive steady states $c,c^*$, it must hold that $\vi/\vii$ $=$ $\viaster/\viiaster$ which can be rewritten as $\vi/\viaster$ $=$ $\vii/\viiaster$.
		This leads to the following definition:
	
		\definition{definition_coarsest_partition_reactions}{statements}{
			Let $\crn$ be a chemical reaction network that admits a positive steady state.	
			Two reactions $\ri,\rii\in\rset$ are said to be in relation $\myrelrsmall$ if for any pair of positive steady states $c,c^*$, $[v(c)]_{\ri}/[v(c^*)]_{\ri} = [v(c)]_{\rii}/[v(c^*)]_{\rii}$.
		}

		The equivalence relation $\myrelrsmall$ induces a partition of $\rset$ into equivalence classes.
		The resulting partition is denoted by $\mypartr$.
		For a given chemical reaction network, the partition $\mypartr$ is an inherent property of the corresponding system of ODEs.
		The problem now is that of determining the partition $\mypartr$, i.e., all pairs of reactions whose flux ratio at a positive steady state is a constant. 
		To address this nontrivial problem, we first observe that for two reactions $\ri, \rii \in \rset$ with $y = \tilde{y}$, i.e., pairs of reactions which use the same substrate complex, it always holds that $\ri \myrelrsmall \rii$, as shown by the following lemma:

		\lemma{lemma_same_substrate}{statements}{
			Let $\crn$ be a chemical reaction network. 
			If two reactions $\ri, \rii \in\rset$ share the same substrate complex, i.e., $\yi = \yii$ , then $\ri \myrelrsmall \rii$.\bigskip 
		}{
			The reactions $y\rightarrow y',y\rightarrow \tilde{y}'\in\mathcal{R}$ share the same substrate complex.
			It follows directly that
			\begin{equation*}
				\frac{\left[v(c)\right]_{y\rightarrow y'}}{\left[v(c^*)\right]_{y\rightarrow y'}} = 
				\frac{k_{y\rightarrow y'} c^y}{k_{y\rightarrow y'} c^{*y}} = 
				\frac{c^y}{c^{*y}} =
				\frac{k_{y\rightarrow \tilde{y}'} c^y}{k_{y\rightarrow \tilde{y}'} c^{*y}} = 
				\frac{\left[v(c)\right]_{y\rightarrow \tilde{y}'}}{\left[v(c^*)\right]_{y\rightarrow \tilde{y}'}},
			\end{equation*}
			and, therefore, $y\rightarrow y'\myrelrsmall\tilde{y}\rightarrow\tilde{y}'$.
		}

		Consider the partition of the set of reactions based on whether the reactions of the same equivalence class use the same substrate complex.
		Let the corresponding partition of the set of reactions be denoted by $\rset/\myrelrsmallx{0}$.
		Lemma \ref{lemma_same_substrate} shows that reactions, belonging to the same equivalence class in $\rset/\myrelrsmallx{0}$, always exhibit the same reaction rate ratio.
		Moreover, if two reactions use the same substrate complex, then they are element of the same equivalence class in $\rset/\myrelrsmall$.
		Then, for each $\dset^0\in\rset/\myrelrsmallx{0}$ there exists $\dset\in\rset/\myrelrsmall$ with $\dset^0\subseteq\dset$, and it follows that $\rset/\myrelrsmallx{0}\leq\rset/\myrelrsmall$.
		The partitions of the set of reactions which are not finer than $\rset/\myrelrsmallx{0}$ and not coarser than $\rset/\myrelrsmall$ belong to the following set:

		\definition{definition_P_R}{statements}{
			Let $\crn$ be a chemical reaction network that admits a positive steady state.
			Then,
			\begin{equation*}
				\pset_{\rset} := \{\rset/\myrelrsmallx{'} \myspace\vert\myspace \rset/\myrelrsmallx{0}\leq\rset/\myrelrsmallx{'}\myspace\text{and}\myspace\rset/\myrelrsmallx{'}\leq\mypartr\}.
			\end{equation*}\vspace{-0.5cm}
		}
		
		Clearly, the partition $\mathcal{R}/\myrelrsmallx{0}$ represents the finest element in $\pset_{\rset}$.
		Moreover, it is also the finest partition of the set of reactions which is compatible with mass action kinetics, i.e., a partition such that all reactions with substrate complex $y$ are in the same equivalence class.
		Analogously, the partition $\mathcal{R}/\myrelrsmall$ represents the coarsest element in $\mathcal{P}_{\mathcal{R}}$.
		If two reactions $y\rightarrow y',\tilde{y}\rightarrow\tilde{y}'\in\mathcal{R}$ are in the equivalence relation $\myrelrsmallx{'}$ with $\mathcal{R}/\myrelrsmallx{'}\in\mathcal{P}_{\mathcal{R}}$, then it directly follows that $[v(c)]_{y\rightarrow y'}/[v(c^*)]_{y\rightarrow y'}$ $=$ $[v(c)]_{\tilde{y}\rightarrow\tilde{y}'}/[v(c^*)]_{\tilde{y}\rightarrow\tilde{y}'}$ and, therefore, $\ri\myrelrsmall\rii$.
		The set $\pset_{\rset}$, together with the relation $\leq$ on its elements, as given in Definition \ref{definition_partition_less_or_equal}, is a partially ordered set (see Lemma S\ref{supp_lemma_P_R_is_poset} in the Appendix for the proof).
		Furthermore, together with the binary join and meet, from Definitions S\ref{supp_definition_binary_operation_meet} and S\ref{supp_definition_binary_operation_join} in the Appendix, $\pset_{\rset}$ represents a lattice (see Example \ref{example_example_partitions} and Corollary S\ref{supp_corollary_P_R_lattice} in the Appendix for the proof).
		In contrast to the partition $\rset/\myrelrsmall$, the partition $\rset/\myrelrsmallx{0}$ can easily be determined by investigating the substrate complexes of the set of reactions.
		As a result, the higher a given element from $\pset_{\rset}$ is located in the lattice, the more information about invariant reaction ratios becomes available.

		\example{example_example_partitions}{examples}{
			The lower bound of the lattice corresponding to Example \ref{example_chemical_reaction_network} is given by
			\begin{equation*}
				\rset/\myrelrsmallx{0} = 
					\mybig{\{}
						\{R_1\},
						\{R_2,R_4\},
						\{R_3\},
						\{R_5\},
						\{R_6\},
						\{R_7\},
						\{R_8\}
					\mybig{\}},
			\end{equation*}
			and, provided an oracle that yields $\rset/\myrelrsmall$, the upper bound is given by
			\begin{equation*}
				\rset/\myrelrsmall =
					\mybig{\{}
						\{R_1,R_2,R_3,R_4\},
						\{R_5,R_6,R_7,R_8\}
					\mybig{\}}.
			\end{equation*}
			Then, $\pset_{\rset}$ consists of $\rset/\myrelrsmallx{0}$ and $\rset/\myrelrsmall$ as well as all partitions of $\rset$ that are coarser than $\rset/\myrelrsmallx{0}$ but finer than $\rset/\myrelrsmall$.

			(See the following examples for the derivation of this partition.
			For $k_{R_1} = 1$, $k_{R_2} = 1$, $k_{R_3} = 1$, $k_{R_4} = 1$, $k_{R_5} = 1$, $k_{R_6} = 1$, $k_{R_7} = 1$, $k_{R_8} = 1$, there exist at least two positive steady states: (i) $\myconc{c}{A} = 1$, $\myconc{c}{B} = 1$, $\myconc{c}{C} = 1$, $\myconc{c}{D} = 1$, $\myconc{c}{E} = 1$, $\myconc{c}{X} = 1$; and (ii) $\myconc{c}{A} = 7/6$, $\myconc{c}{B} = 7/6$, $\myconc{c}{C} = 3/2$, $\myconc{c}{D} = 7/4$, $\myconc{c}{E} = 7/4$, $\myconc{c}{X} = 7/6$; which implies that the rate ratio of $R_1$ and $R_8$ is not constant for each positive steady state.
			Therefore, the partition consisting of two equivalence classes is in fact the coarsest.)
		}
		
		One obvious question arises:
		What aspects of the chemical reaction network can be used to coarsen the finest partition in $\pset_{\rset}$?
		It turns out that the kernel of the stoichiometric matrix can directly be used to answer this question, as illustrated in Example \ref{example_kernel} (see also the concept of co-sets in \citet{Papin2004}).
		That is, the computation of fully coupled reactions from FCA using a linear program is not necessary.
		
		\example{example_kernel}{examples}{
			A rational basis of the kernel of the stoichiometric matrix of Example \ref{example_reactions} (see Example \ref{example_stoichiometric_matrix} for matrix $N$) is given by
			\begin{equation*}
				\myarraythree{\ker(N) =}{x{\wsmall}x{\bw}x{\bw}x{\bw}}{
					1 & 0 & 0 & 0\tabularnewline
					1 & 0 & 0 & 0\tabularnewline
					0 & 1 & 0 & 0\tabularnewline
					0 & 1 & 0 & 0\tabularnewline
					0 & 0 & 1 & 0\tabularnewline
					0 & 0 & 1 & 0\tabularnewline
					0 & 0 & 0 & 1\tabularnewline
					0 & 0 & 0 & 1
				}{
 					R_1\\
 					R_2\\
 					R_3\\
 					R_4\\
 					R_5\\
 					R_6\\
 					R_7\\
 					R_8
				},
			\end{equation*}
			which implies the following partition of the set of reactions
			\begin{equation*}
				\rset/\thicksim = 
					\mybig{\{}
						\{R_1,R_2\},
						\{R_3,R_4\},
						\{R_5,R_6\},
						\{R_7,R_8\}
					\mybig{\}}.
			\end{equation*}
			Since reactions $R_2$ and $R_4$ are in the relation $\myrelrsmallx{0}$, this implies that reactions $R_1$, $R_2$, $R_3$ and $R_4$ satisfy Definition \ref{definition_coarsest_partition_reactions}.
			Therefore, one obtains
			\begin{equation*}
				\rset/\myrelrsmallx{''} =
					\mybig{\{}
						\{R_1,R_2,R_3, R_4\},
						\{R_5,R_6\},
						\{R_7,R_8\}
					\mybig{\}},
			\end{equation*}
			with $\rset/\myrelrsmallx{''} \in\pset_{\rset}$.
		}

	\subsection{Invariant complex ratios}
		Analogous to the previous subsection, consider two complexes $\yi,\yii\in\cset$ such that, at each positive steady state $c$, the ratio $c^{\yi}/c^{\yii}$ is invariant.
		Then, for two positive steady states $c ,c^*$, it must hold that $c^{\yi}/c^{\yii} = c^{*\yi}/c^{*\yii}$, which can be rewritten as $c^{\yi}/c^{*\yi} = c^{\yii}/c^{*\yii}$.
		This leads to the following definition:

		\definition{definition_coarsest_partition_complexes}{statements}{
			Let $\crn$ be a chemical reaction network that admits a positive steady state.
			Two complexes $y,y'\in\cset$ are said to be in relation $\myrelcsmall$ if for every pair of positive steady states $c,c^*$, $c^y/c^{*y} = c^{\tilde{y}}/c^{*\tilde{y}}$.
		}
		
		The equivalence relation $\myrelcsmall$ induces a partition of $\cset$ into equivalence classes.
		The resulting partition is denoted by $\mypartc$.
		Let the partition $\{\{y\}\myspace\vert\myspace y\in\cset\}$ of the set of complexes be denoted by $\cset/\myrelcsmallx{0}$.
		Trivially, pairs of elements from an equivalence class in $\cset/\myrelcsmallx{0}$ are, according to Definition \ref{definition_coarsest_partition_complexes}, in the equivalence relation $\myrelc$.
		Furthermore, it is not difficult to see that $\cset/\myrelcsmallx{0}\leq\cset/\myrelcsmall$.
		Consequently, the set of partitions of the set of complexes which are not finer than $\cset/\myrelcsmallx{0}$ and not coarser than $\cset/\myrelcsmall$ can be defined, analogous to $\pset_{\rset}$, as follows:

		\definition{definition_P_C}{statements}{
			Let $\crn$ be a chemical reaction network that admits a positive steady state.
			Then,
			\begin{equation*}
				\pset_{\cset} := \{\cset/\myrelcsmallx{'} \myspace\vert\myspace \cset/\myrelcsmallx{0}\leq\cset/\myrelcsmallx{'}\myspace\text{and}\myspace\cset/\myrelcsmallx{'}\leq\mypartc\}.
			\end{equation*}\vspace{-0.5cm}
		}

		The partition $\cset/\myrelcsmall$ represents the coarsest element in $\pset_{\cset}$.
		Moreover, if two complexes $y,\tilde{y}\in\cset$ are in the equivalence relation $\myrelcsmallx{'}$, with $\cset/\myrelcsmallx{'}\in\pset_{\cset}$, then it directly follows that $c^y/c^{y*} = c^{\tilde{y}}/c^{*\tilde{y}}$ and, therefore, $y\myrelcsmall\tilde{y}$.
		The set $\pset_{\cset}$, together with the relation $\leq$ on its elements is a partially ordered set (see Lemma S\ref{supp_lemma_P_C_is_poset} in the Appendix for the proof).
		Furthermore, together with the binary join and meet, as given in Definitions S\ref{supp_definition_binary_operation_meet_complexes} and S\ref{supp_definition_binary_operation_join_complexes} in the Appendix, $\pset_{\cset}$ represents a lattice (see Corollary S\ref{supp_corollary_P_C_lattice} in the Appendix for the proof).
		In contrast to the partition $\cset/\myrelcsmall$, the partition $\cset/\myrelcsmallx{0}$ can be trivially determined.
		As a result, the higher a given element from $\pset_{\cset}$ is located in the lattice, the more information about invariant complex ratios becomes available.

	\subsection{$\pset_{\rset}$ is homomorphic to $\pset_{\cset}$}
		In this section, it is shown that there exists a map $\varphi: \pset_{\rset} \rightarrow \pset_{\cset}$, which preserves the structure of $\pset_{\rset}$.
		Let $\rset/\myrelrsmallx{'}\in\pset_{\rset}$ and $\dset\in\rset/\myrelrsmallx{'}$.
		Furthermore, let $\uset_{\dset} = \{y\myspace\vert\myspace\ri\in\dset\}$ and let $\overline{\uset}$ represent the set of complexes that do not participate in any reaction as a substrate.

		\definition{definition_map_set_of_reactions_to_set_of_complexes}{statements}{
			Let $\crn$ be a chemical reaction network that admits a positive steady state.
			Further, suppose that some partition $\rset/\myrelrsmallx{'}\in\pset_{\rset}$ is given.
			The map $\varphi: \pset_{\rset} \rightarrow \pset_{\cset}$ is defined by
			\begin{equation*}
				\varphi(\rset/\myrelrsmallx{'}) := \{\uset_{\dset} \myspace\vert\myspace \dset\in\rset/\myrelrsmallx{'}\}\cup\{\{y\}\myspace\vert\myspace y\in\overline{\uset}\}.
			\end{equation*}\vspace{-0.75cm}
		}

		In fact, one can show that $\pset_{\rset}$ is homomorphic to $\pset_{\cset}$ with respect to $\varphi$ (see Corollary S\ref{supp_phi_homomorphism} in the Appendix for the proof).
		Example \ref{example_map} illustrates the result of applying the map $\varphi$ on the partition from Example \ref{example_kernel}.

		\example{example_map}{examples}{
			Partition $\rset/\myrelrsmallx{''} = \{\{R_1,$ $R_2,$ $R_3,$ $R_4\},$ $\{R_5,$ $R_6\},$ $\{R_7,$ $R_8\}\}$ from Example \ref{example_kernel} can be mapped to a partition of the set of complexes:
			From Definition \ref{definition_map_set_of_reactions_to_set_of_complexes}, it follows that 
			\begin{equation*}
				\varphi(\rset/\myrelrsmallx{''}) =
					\mybig{\{}
						\uset_{\{R_1, R_2, R_3, R_4\}},
						\uset_{\{R_5, R_6\}},
						\uset_{\{R_7, R_8\}}
					\mybig{\}},
			\end{equation*}
			with $\uset_{\{R_1, R_2, R_3, R_4\}} = \uset_{\{\text{A}\rightarrow\text{X}, \text{X}\rightarrow\text{A}, \text{B}\rightarrow\text{X}, \text{X}\rightarrow\text{B}\}} = \{\{\text{A}\},\{\text{B}\},\{\text{X}\}\}$, $\uset_{\{R_5, R_6\}} = \uset_{\{\text{A}+\text{C}\rightarrow\text{D}, \text{D}\rightarrow\text{A}+\text{C}\}} = \{\{\text{A},\text{C}\},\{\text{D}\}\}$ and $\uset_{\{R_7, R_8\}} = \uset_{\{\text{B}+\text{C}\rightarrow\text{E}, \text{E}\rightarrow\text{B}+\text{C}\}} = \{\{\text{B},\text{C}\},\{\text{E}\}\}$.
			Therefore,
			\begin{equation*}
				\varphi(\rset/\myrelrsmallx{''}) =
					\mybigg{\{}
						\mybig{\{}
							\{\text{A}\},\{\text{B}\},\{\text{X}\}
						\mybig{\}}, 
						\mybig{\{}
							\{\text{A},\text{C}\},\{\text{D}\}
						\mybig{\}},
						\mybig{\{}
							\{\text{B},\text{C}\},\{\text{E}\}
						\mybig{\}}
					\mybigg{\}}.
			\end{equation*}\vspace{-0.5cm}
		}

		Now consider two positive steady states $c,c^*$ and a partition of the set of complexes $\cset/\myrelcsmallx{'}\in\pset_{\cset}$, for instance derived by $\varphi$ from a partition $\rset/\myrelrsmallx{'}\in\pset_{\rset}$.
		Since $y,\tilde{y}\in\uset$ with $\uset\in\cset/\myrelcsmallx{'}$ satisfy $c^y/c^{*y} = c^{\tilde{y}}/c^{*\tilde{y}}$, they also satisfy $\ln(c^y/c^{*y}) =\ln(c^{\tilde{y}}/c^{*\tilde{y}})$ and, thus, $(y - \tilde{y})\ln(c/c^*) = 0$.
		It is not difficult to see that $\ln(c/c^*)$ is orthogonal to any element in $\spn(\{y - \tilde{y}\myspace\vert\myspace y,\tilde{y}\in\uset\})$.

		\lemma{lemma_complex_difference}{statements}{
			Let $\crn$ be a chemical reaction network that admits a positive steady state.
			Furthermore, consider a partition of the set of complexes $\cset/\myrelcsmallx{'}\in\pset_{\cset}$.
			If there exist $y,\tilde{y}\in\cset$ such that $y - \tilde{y}\in\spn(\{y' - \tilde{y}'\myspace\vert\myspace y',\tilde{y}'\in\uset\})$ with $\uset\in\cset/\myrelcsmallx{'}$, then $y \myrelcsmall \tilde{y}$.
		}{
			Let $c,c^*$ be two positive steady states of the chemical reaction network.
			The vector $\ln(c/c^*)$ is orthogonal to each vector in $\spn(\{y' - \tilde{y}'\myspace\vert\myspace y',\tilde{y}'\in\uset\})$ and, thus, $\ln(c/c^*)$ is also orthogonal to $y - \tilde{y}$.
			It follows that $(y - \tilde{y})\ln(c/c^*) = 0$ and, therefore, $y \myrelcsmall \tilde{y}$.
		}

		Here we note that Lemma \ref{lemma_complex_difference} is very similar to Proposition S$4.1$ in \citet{Shinar2010} (Supporting Information).
		Shinar and Feinberg use their proposition to investigate invariance of concentrations in chemical reaction networks.
		However, with our lemma we attempt to coarsen the partition of the set of complexes.
		We show that the knowledge of the coarsest partition reduces the complexity of the system since dependencies between reactions and complexes become visible.
		In fact, by following the concentration-centered view of CRNT, as pursued in \citet{Shinar2010}, one may neglect the information contained in the dependence between reactions and complexes.
		Example \ref{example_complex_decomposition} illustrates the implications of Lemma \ref{lemma_complex_difference}.

		\example{example_complex_decomposition}{examples}{
			Lemma \ref{lemma_complex_difference} applies to the partition $\cset/\myrelcsmallx{''} = \varphi(\rset/\myrelrsmallx{''})$ from Example \ref{example_map} as follows:
			$\uset = \{\{\text{A}\},\{\text{B}\},\{\text{X}\}\}\in\cset/\myrelcsmallx{''}$ and $y = \{\text{A},\text{C}\}, y' = \{\text{B},\text{C}\}\in\cset$.
			Then, $y - \tilde{y}\in\spn(\{y' - \tilde{y}'\myspace\vert\myspace y',\tilde{y}'\in\uset\})$ since there exist $y',\tilde{y}'\in\uset$ with $y'-\tilde{y}' = \{\text{A}\} - \{\text{B}\} = \{\text{A},\text{C}\} - \{\text{B},\text{C}\} = y - y'$.
			This results in the following, coarser partition of the complexes
			\begin{equation*}
				\cset/\myrelcsmallx{'} =
					\mybigg{\{}
						\mybig{\{}
							\{\text{A}\},\{\text{B}\},\{\text{X}\}
						\mybig{\}}, 
						\mybig{\{}
							\{\text{A},\text{C}\},\{\text{D}\},\{\text{B},\text{C}\},\{\text{E}\}
						\mybig{\}}
					\mybigg{\}}.
			\end{equation*}\vspace{-0.5cm}			
		}
	
		Let $\dset_{\uset} := \{\ri\myvert y\in\uset\myand\ri\in\rset\}$ be the set of reactions which have the complexes in $\uset$ as substrates.
		We next need the following definition which maps a partition of the set of complexes back to a partition of the set of reactions:
	
		\definition{definition_mapback}{statements}{
			Let $\crn$ be a chemical reaction network that admits a positive steady state.
			Furthermore, consider a partition $\cset/\myrelcsmallx{'}\in\pset_{\cset}$.
			The map $\mu: \pset_{\cset} \rightarrow \pset_{\rset}$ is defined by
			\begin{equation*}
				\mu(\cset/\myrelcsmallx{'}) := \{\dset_{\uset} \myspace\vert\myspace \uset\in\cset/\myrelcsmallx{'}\}\setminus\emptyset.
			\end{equation*}\vspace{-0.75cm}
		}
	
		Application of the map $\mu$ to a partition of the set of complexes is illustrated in Example \ref{example_coarsest_partition}.
	
		\example{example_coarsest_partition}{examples}{
			Partition $\cset/\myrelcsmallx{'} = \{\{\{\text{A}\},$ $\{\text{B}\},$ $\{\text{X}\}\},$ $\{\{\text{A},$ $\text{C}\},$ $\{\text{D}\},$ $\{\text{B},$ $\text{C}\},$ $\{\text{E}\}\}\}$ from Example \ref{example_complex_decomposition} can be mapped to a partition of the set of reactions:
			From Definition \ref{definition_mapback}, it follows that 
			\begin{equation*}
				\mu(\cset/\myrelcsmallx{'}) =
					\mybig{\{}
						\dset_{\{\{\text{A}\},\{\text{B}\},\{\text{X}\}\}},
						\dset_{\{\{\text{A},\text{C}\},\{\text{D}\},\{\text{B},\text{C}\},\{\text{E}\}\}}
					\mybig{\}},
			\end{equation*}
			with $\dset_{\{\{\text{A}\},\{\text{B}\},\{\text{X}\}\}} = \{R_1, R_2, R_3, R_4\}$ and $\dset_{\{\{\text{A},\text{C}\},\{\text{D}\},\{\text{B},\text{C}\},\{\text{E}\}\}} = \{R_5, R_6, R_7, R_8\}$.
			Therefore,
			\begin{equation*}
				\mu(\cset/\myrelcsmallx{'}) = 
				\mybig{\{}
					\{R_1, R_2, R_3, R_4\},
					\{R_5, R_6, R_7, R_8\}
				\mybig{\}}.
			\end{equation*}\vspace{-0.5cm}
		}
		
		It should be apparent that there may exist partitions in $\pset_{\cset}$ that are coarser than $\varphi(\rset/\myrelrsmall)$.
		Such partitions can consist of equivalence classes $\uset$ with $\vert\uset\vert\geq 2$ which also contain complexes from $\overline{\uset}$.
		This information is lost when mapping such a partition of the set of complexes to a partition of the set of reactions using $\mu$.
		As a result, $\varphi$ is an injective map while $\mu$ is surjective.

		Consequently, the map $\varphi$ has the useful property that a coarsening of a partition of the set of reactions implies a coarsening of the corresponding partition of the set of complexes.
		Furthermore, the map $\mu$ has the property that a coarsening of the partition of the set of complexes guarantees that the corresponding partition of the set of reactions is not refined.		
					
	\subsection{Connections to CRNT}
		\citet{Shinar2010} defined an equivalence relation on the complexes, denoted by $\leftrightsquigarrow$, as follows:
		two complexes $y,\tilde{y}\in\mathcal{C}$ are in the relation $\leftrightsquigarrow$ if, for two positive steady states $c,c^*$, $(y - \tilde{y})\ln(c/c^*) = 0$ is satisfied.
		This equivalence relation induces a partition of the complexes.
		It is not difficult to see that this condition is equivalent to our Definition \ref{definition_coarsest_partition_complexes}.
		
		Interestingly, \citet{Shinar2010} specified which complexes are in the relation given by Definition \ref{definition_coarsest_partition_complexes} only for special classes of networks.
		CRNT defines several properties of reaction networks, e.g., the deficiency and weakly reversibility \citep{Feinberg1979, Feinberg1995, Gunawardena2003}.
		The deficiency $\delta$ is an index which can provide information about the dynamic behavior of a mass action system independently of the rate constants.
		This index is computed as $\delta = n - l - q$, where $n$ is the number of complexes, $l$ is the number of linkage classes, i.e., the number of connected components of the graph which can be build by the reactions (see Example \ref{example_crnt_vs_us} for an illustration), and $q$ is the rank of the stoichiometric matrix $N$.
		A reaction network is said to be weakly reversible, if, in each connected component, there exists a path from any node (i.e., complex) to all other nodes in the connected component.
		For instance, in deficiency-zero reaction networks which are weakly reversible, complexes are in the same equivalence class if they are elements of the same linkage class \citep{Shinar2010}.
		Furthermore, in deficiency-one reaction networks, all complexes of nonterminal strong linkage classes are elements of the same equivalence class \citep{Shinar2010}.
		With our definition of the lattice $\pset_{\cset}$, it is clear that in both cases the resulting partitions are elements of $\pset_{\cset}$.
		Nevertheless, they may still not represent the coarsest element. 

		In contrast to the results from CRNT, our findings do not pertain to special classes of networks.
		In fact, our theoretical results deal with equivalence classes of reactions which can be mapped to a partition of the set of complexes in general networks.
		The defined map in turn enables the determination of coarser partitions of the set of complexes from coarser partitions of the set of reactions.

		\example{example_crnt_vs_us}{examples}{
			The chemical reaction network from Example \ref{example_chemical_reaction_network} can be represented by the following graph:
			
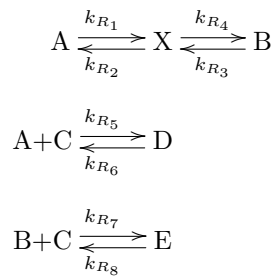
\begin{equation*}
				\begin{split}
					\xymatrix{
						\text{\ \ \ \ A} \ar@<0.5ex>[r]^{k_{R_1}} & \ar@<0.5ex>[l]^{k_{R_2}} \text{X} \ar@<0.5ex>[r]^{k_{R_4}} & \ar@<0.5ex>[l]^{k_{R_3}} \text{B}\\
						\text{A+C} \ar@<0.5ex>[r]^{k_{R_5}}                		   & \ar@<0.5ex>[l]^{k_{R_6}} \text{D} & \\
			     		\text{B+C} \ar@<0.5ex>[r]^{k_{R_7}}       		           & \ar@<0.5ex>[l]^{k_{R_8}} \text{E} &
					}
				\end{split}
			\end{equation*}
			The set of nodes of this graph is equivalent to the set of complexes (here $n = 7$).
			The connected components of this graph are equivalent to the linkage classes (here $l = 3$).
			The strongly connected components of this graph are equivalent to the strong linkage classes (here equivalent to the linkage classes).
			The terminal strong linkage classes are the strong linkage classes from which there exists no path to another strong linkage class (here equivalent to the linkage classes).
			Thus, this reaction network is weakly reversible.
			The rank of the stoichiometric matrix $N$ is $q = 4$ (see also Example \ref{example_stoichiometric_matrix}).
			As a result, the deficiency of this reaction network is $\delta = n - l - q = 0$.
			Consequently, the coarsest partition of the set of complexes with respect to Definition \ref{definition_coarsest_partition_complexes} contains at most three equivalence classes.
			However, by applying our approach, there exists a coarser partition consisting of two equivalence classes, as illustrated in Example \ref{example_coarsest_partition}.
		}
		
\section{Conclusions}
We analyzed ratio invariants in chemical reaction networks on the level of reactions and on the level of complexes.
We show that there can exist pairs of distinct reactions whose reaction rate ratio is constant in each steady state.
This fact can be used to define a partition of the set of reactions.
The knowledge of all such pairs of reactions determines the coarsest partition of the set of reactions ($\rset/\myrelrsmall$).
Furthermore, we defined the set of partitions which are compatible with mass action kinetics ($\pset_{\rset}$), i.e., reactions which use the same substrate complex must be elements of the same equivalence class.
As a result, the coarsest partition of the set of reactions cannot be refined arbitrarily without violating constraints imposed by mass action kinetics.
The idea of existence of the coarsest and finest partition ($\rset/\myrelrsmallx{0}$) of the set of reactions, led to the introduction of the lattice of partitions of the set of reactions, represented by $\pset_{\rset}$.

Analogously, we defined a finest ($\cset/\myrelcsmallx{0}$) and a coarsest partition ($\cset/\myrelcsmall$) on the set of complexes as well as the corresponding lattice $\pset_{\cset}$.
Since the rate of a given reaction is directly influenced by the product of the concentrations of its substrates, we introduced an injective map ($\varphi$) which converts a partition of the set of reactions to a partition of the set of complexes.
Additionally, we defined a surjective map ($\mu$) which converts a partition of the set of complexes to a partition of the set of reactions.

We explicitly point out that the coarsest partition of the set of reactions, and, thus, of the complexes, might depend on the rate constants.
As illustrated by the examples, there always exist partitions of the set of reactions which can be derived independently of the rate constants.
However, at this point, it is not clear whether the coarsest partition of the set of reactions changes for different choices of the set of rate constants.

The study of metabolic networks has been hampered by the dichotomy in the computational approaches focused either on analysis of biochemical network fluxes or on concentrations of biochemical network components.
Our study bridges this dichotomy by establishing the relation between the flux-centric and concentration-centric approaches, here represented by FCA and CRNT.
Based on the established connection between invariants on the level of reactions and on the level of concentrations, we provide a method that will allow a deeper algebraic insight into the dynamic behavior of chemical mass action systems avoiding numerical computations.
Our theoretical findings also provide the impetus for rigorous analysis of biological systems concerning the dynamics of biochemical networks.

\section{Acknowledgments}
JN and ZN are supported by the Max-Planck Society.
SG is supported by the ColoNET project funded by the Federal Ministry of Education and Research, Grant no. 0315417F.

\setcounter{statements}{0}
\newpage
\appendix
\section{Appendix}

\subsection{Prerequisites}
	The following general definitions are indispensable to understand the notation we use in this work: 

	\definitionsupplement{supp_definition_equivalence_relation}{statements}{
		A relation $\thicksim$ on a set $\aset$ is called an \defbf{equivalence relation}, if it satisfies
		\begin{itemize}
			\item[  (i)] $x \thicksim x\myspace\vert\myspace\forall x\in\aset$ (reflexivity),
			\item[ (ii)] if $x,y\in\aset$ and $x \thicksim y$, then $y \thicksim x$ (symmetry), and
			\item[(iii)] if $x,y,z\in\aset$, $x\thicksim y$ and $y\thicksim z$, then $x\thicksim z$ (transitivity).
		\end{itemize}
	}\vspace{-0.3cm}

	\definitionsupplement{supp_definition_equivalence_class}{statements}{
		Consider a relation $\thicksim$ on a set $\aset$ and an element $x\in\aset$.
		The set of elements $y\in\aset$ with $x\thicksim y$ is called the \defbf{equivalence class} of $x$.
		The set of equivalence classes of $\aset$ is denoted by $\aset/\thicksim$.
	}

	\definitionsupplement{supp_definition_partition}{statements}{
		Let $\aset$ be a set and let $\dset_1, \ldots, \dset_n \subseteq \aset$.
		The set $\{\dset_1, \ldots, \dset_n\}$ is called a \defbf{partition} of $\aset$ if and only if
		\begin{itemize}
			\item[  (i)] $\dset_1 \cup \ldots \cup \dset_n = \aset$ and
			\item[ (ii)] $\dset_i \cap \dset_j = \emptyset$ with $\dset_i,\dset_j\in\aset$ and $\dset_i\neq \dset_j$.
		\end{itemize}\vspace{-0.3cm}
	}

	The equivalence classes of a set $\mathcal{A}$ with respect to a relation $\thicksim$ yield a partition of $\mathcal{A}$ \citep{Makinson2008}.
	Therefore, we use the terms ``equivalence classes'' and ``partition'' of a set with respect to a given relation synonymously.\\

	Given a chemical reaction network $\crn$ and a partition $\rset/\thicksim$ of the set of reactions, we define $\uset_{\dset} := \{y\myspace\vert\myspace\ri\in\dset\}$ for $\dset\in\rset/\thicksim$.
	Furthermore, let
	\begin{equation*}
		\overline{\uset} := \cset\setminus\bigcup_{\dset\in\rset/\thicksim}\uset_{\dset}
	\end{equation*}
	denote the set of complexes that do not participate in any reaction as substrates.
	Additionally, let $\overline{\yset} := \{\{y\}\myvert y\in\overline{\uset}\}$.
	Analogously, given a partition $\cset/\thicksim$ of the corresponding set of complexes, we define $\dset_{\uset} := \{\ri\myvert y\in\uset\myand\ri\in\rset\}$ for $\uset\in\cset/\thicksim$.

	For graph-theoretical concepts, see \citet{Bollobas1998}.
						
\subsection{$\pset_{\rset}$ is a lattice}
		The set $\pset_{\rset}$ is defined as follows (see the main text for the definition of $\rset/\myrelrsmallx{0}$ and $\mypartr$):

		\definitionsupplement{supp_definition_P_R}{statements}{
			Let $\crn$ be a chemical reaction network that admits a positive steady state.
			Then,
			\begin{equation*}
				\pset_{\rset} := \{\rset/\myrelrsmallx{'} \myspace\vert\myspace \rset/\myrelrsmallx{0}\leq\rset/\myrelrsmallx{'}\myspace\text{and}\myspace\rset/\myrelrsmallx{'}\leq\mypartr\}.
			\end{equation*}\vspace{-0.5cm}
		}

		Then, $\pset_{\rset}$ is a partially ordered set as the following lemma shows:

		\lemmasupplement{supp_lemma_P_R_is_poset}{statements}{
			Let $\crn$ be a chemical reaction network that admits a positive steady state.
			The set $\mathcal{P}_{\rset}$ is a partially ordered set.
		}{
			For $\mathcal{A}, \mathcal{A}', \mathcal{A}'' \in \mathcal{P}_{\rset}$, it holds that:
			\begin{itemize}
			  	\item $\mathcal{A} \leq \mathcal{A}$ by definition.
				\item If $\mathcal{A} \leq \mathcal{A}'$ and $\mathcal{A}' \leq \mathcal{A}$, then $\forall \dset\in\mathcal{A}$, $\exists \dset'\in\mathcal{A}'$ with $\dset \subseteq \dset'$ and $\forall \dset'\in\mathcal{A}'$, $\exists \dset''\in\mathcal{A}$ with $\dset' \subseteq \dset''$ which implies $\dset\subseteq\dset''$.
					  Since $\dset,\dset''\in\aset$, it follows that $\dset = \dset''$, implying $\dset \subseteq \dset'$ and $\dset' \subseteq \dset$.
					  Therefore, $\aset = \aset'$.
				\item If $\mathcal{A} \leq \mathcal{A}'$ and $\mathcal{A}' \leq \mathcal{A}''$, then $\forall \dset\in\mathcal{A}$, $\exists \dset'\in\mathcal{A}'$ with $\dset \subseteq \dset'$ and $\forall \dset'\in\mathcal{A}'\myspace\exists \dset''\in\mathcal{A}''$ with $\dset' \subseteq \dset''$.
				      It follows directly that $\forall \dset\in\mathcal{A}\myspace\exists \dset''\in\mathcal{A}''$ with $\dset \subseteq \dset''$ and, therefore, $\mathcal{A} \leq \mathcal{A}''$.
			\end{itemize}
			Therefore, $\mathcal{P}_{\rset}$ is a partially ordered set.
		}

		The following argument will be crucial in what follows:
		Let $\aset,\aset',\aset''$ be partitions of $\rset$ and let $\aset\leq\aset''$ and $\aset'\leq\aset''$.
		Furthermore, let $\dset\in\aset$ and $\dset'\in\aset'$.
		From the properties of a partial order (see Lemma S\ref{supp_lemma_P_R_is_poset}), w.l.o.g., it follows that $\exists \dset_1'',\dset_2''\in\aset''$ with $\dset\subseteq\dset_1''$ and $\dset'\subseteq\dset_2''$.
		If $\dset\cap\dset'\neq\emptyset$, then $\dset_1'' = \dset_2''$, since $\aset''$ is a partition of $\rset$, implying $\dset\cup\dset'\subseteq\dset_1''$.

		\definitionsupplement{supp_definition_binary_operation_meet}{statements}{
			Let $\crn$ be a chemical reaction network that admits a positive steady state.
			The binary operation $\mathcal{A} \wedge\mathcal{A}'$ of two elements $\mathcal{A},\mathcal{A}'\in\mathcal{P}_{\mathcal{R}}$ is defined by $\mathcal{A} \wedge\mathcal{A}' := \{\dset \cap \dset' \myspace\vert\myspace \dset\in\mathcal{A}\myspace\text{and}\myspace \dset'\in\mathcal{A}'\myspace\text{and}\myspace \dset\cap\dset' \neq\emptyset\}$.
		}

		\lemmasupplement{supp_lemma_meet_is_partition}{statements}{
			Let $\crn$ be a chemical reaction network that admits a positive steady state.
			Given two elements $\mathcal{A},\mathcal{A}'\in\mathcal{P}_{\mathcal{R}}$, the set $\mathcal{A} \wedge\mathcal{A}'$ is a partition of the set of reactions.
		}{
			For each $\ri\in\rset$, there exist exactly one $\dset\in\mathcal{A}$ and exactly one $\dset'\in\mathcal{A}'$ with $\ri\in\dset$ and $\ri\in\dset'$, since $\mathcal{A}$ and $\mathcal{A}'$ are partitions of the set of reactions, so that $\ri\in\dset\cap\dset'\neq\emptyset$.
			It follows directly that for each $\ri\in\rset$ there exists exactly one $\dset\in\mathcal{A}\wedge\mathcal{A}'$ with $\ri\in\dset$.
			As a result, the union of all elements in $\mathcal{A}\wedge\mathcal{A}'$ equals $\rset$ and $\dset\cap\dset' = \emptyset$ for $\dset,\dset'\in\mathcal{A}\wedge\mathcal{A}'$ and $\dset\neq\dset'$.
			Therefore, $\mathcal{A}\wedge\mathcal{A}'$ is a partition of the set of reactions.
		}

		\lemmasupplement{supp_lemma_meet_in_P_R}{statements}{
			Let $\crn$ be a chemical reaction network that admits a positive steady state.
			Given two elements $\mathcal{A},\mathcal{A}'\in\mathcal{P}_{\mathcal{R}}$, the set $\mathcal{A} \wedge\mathcal{A}'$ is an element of $\pset_{\rset}$.
		}{
			If $\ri\in\dset$ with $\dset\in\aset$, then $\dset_{\{y\}}\subseteq\dset$.
			It follows that for each $\dset_{\{y\}}\subseteq\dset$ there exists exactly one $\dset'\in\aset'$ with $\dset_{\{y\}}\subseteq\dset\cap\dset'$.
			As a result, for each $\dset_{\{y\}}$ with $y\in\cset\setminus\overline{\uset}$ there exists exactly one $\dset''\in\mathcal{A}\wedge\mathcal{A}'$ with $\dset_{\{y\}}\subseteq\dset''$.
			Then, $\mathcal{A}\wedge\mathcal{A}'$ is at least as coarse as $\{\dset_{\{y\}}\myspace\vert\myspace y\in\cset\setminus\overline{\uset}\}$ which is in turn the finest element in $\pset_{\rset}$.

			Lemma S\ref{supp_lemma_meet_is_partition} shows that $\aset\wedge\aset'$ is a partition of $\rset$.
			From Definition S\ref{supp_definition_binary_operation_meet}, it follows that $\forall\dset''\in\aset\wedge\aset'$, $\exists\dset\in\aset$ with $\dset''\subseteq\dset$.
			Analogously, $\forall\dset''\in\aset\wedge\aset'$, $\exists\dset'\in\aset'$ with $\dset''\subseteq\dset'$.
			Then, it also follows $\aset\wedge\aset'\leq\aset$ and $\aset\wedge\aset'\leq\aset'$.
			Therefore, $\aset\wedge\aset'\in\pset_{\rset}$.
		}

		\lemmasupplement{supp_lemma_meet_is_greatest_smaller}{statements}{
			Let $\crn$ be a chemical reaction network that admits a positive steady state.
			If $\mathcal{A},\mathcal{A}'\in\pset_{\rset}$, then $\mathcal{A}\wedge\mathcal{A}'$ is the greatest element in $\pset_{\rset}$ which is finer than or equal to $\mathcal{A}$ and $\mathcal{A}'$.
			Therefore, the binary operation $\mathcal{A}\wedge\mathcal{A}'$ is the \defbf{meet} of $\aset, \aset'$.
		}{
			The statement is proved by contradiction:
			Suppose that $\exists\aset''$ with $\aset\wedge\aset'\leq\aset''$, $\aset''\leq\aset$ and $\aset''\leq\aset'$.
			Further suppose that $\aset\wedge\aset'\neq\aset''$.
			Then, $\exists\dset\in\aset$, $\dset'\in\aset'$ and $\dset''\in\aset''$ with $\emptyset\neq\dset\cap\dset'\in\aset\wedge\aset'$ and $\dset\cap\dset'\subset\dset''$.
			From $\aset''\leq\aset$ and $\aset''\leq\aset'$ it also follows that $\dset''\subseteq\dset$ and $\dset''\subseteq\dset'$.
			But this directly implies $\dset''\subseteq\dset\cap\dset'$ which is a contradiction to $\dset\cap\dset'\subset\dset''$.
		}

		\definitionsupplement{supp_definition_binary_operation_join}{statements}{
			Let $\crn$ be a chemical reaction network that admits a positive steady state and let $\aset,\aset'\in\pset_{\rset}$.
			Furthermore, let $G := (\vset, \eset)$ be a graph with $\vset = \{\dset\cup\dset'\myvert\dset\in\aset, \myspace\dset'\in\aset', \myand\dset\cap\dset'\neq\emptyset\}$ and $\eset=\{(v,v')\myvert v,v'\in\vset\myand v\cap v'\neq\emptyset\}$.
			Let $G$ be decomposed as $G = G^{(1)} \cup \ldots \cup G^{(s)}$ where $G^{(i)} = (\vset^{(i)},\eset^{(i)})$ with $1\leq i\leq s$ represent the connected components.
			Then, the binary operation $\aset\vee\aset'$ is defined by
			\begin{equation*}
				\aset\vee\aset':=\Biggl\{\bigcup_{v\in\vset^{(i)}} v\myspace\vert\myspace 1\leq i\leq s\Biggr\}.
			\end{equation*}
			\vspace{-0.3cm}
		}

		\lemmasupplement{supp_lemma_join_is_partition}{statements}{
			Let $\crn$ be a chemical reaction network that admits a positive steady state.
			Given two elements $\mathcal{A},\mathcal{A}'\in\mathcal{P}_{\mathcal{R}}$, the set $\mathcal{A} \vee\mathcal{A}'$ is a partition of the set of reactions.
		}{
			Since two nodes of $G$ are adjacent if the intersection of the corresponding sets is nonempty, it follows that $\ri\in\rset$ is element of exactly one $\dset\in\aset\vee\aset'$.
			As a result, the union of all elements in $\aset\vee\aset'$ equals $\rset$, $\dset\cap\dset' = \emptyset$ for $\dset,\dset'\in\aset\vee\aset'$, and $\dset\neq\dset'$.
			Therefore, $\aset\vee\aset'$ is a partition of the set of reactions.
		}

		\lemmasupplement{supp_lemma_join_in_P_R}{statements}{
			Let $\crn$ be a chemical reaction network that admits a positive steady state.
			Given two elements $\mathcal{A},\mathcal{A}'\in\mathcal{P}_{\mathcal{R}}$, the set $\mathcal{A} \vee\mathcal{A}'$ is an element of $\pset_{\rset}$.
		}{
			Lemma S\ref{supp_lemma_join_is_partition} shows that $\aset\vee\aset'$ is a partition of $\rset$.
			From Definition S\ref{supp_definition_binary_operation_join}, it follows that $\forall\dset\in\aset$, $\exists\dset''\in\aset\vee\aset'$ with $\dset\subseteq\dset''$.
			Analogously, the same holds true for the elements of $\aset'$. 
			Then, it follows that $\aset\leq\aset\vee\aset'$ and $\aset'\leq\aset\vee\aset'$ such that $\aset\vee\aset'$ is as coarse as the finest element in $\pset_{\rset}$.

			Let $\aset''$ be the coarsest element in $\pset_{\rset}$, i.e., $\aset\leq\aset''$ and $\aset'\leq\aset''$.
			If $\dset\cap\dset'\neq\emptyset$ with $\dset\in\aset$ and $\dset'\in\aset'$, then $\exists\dset''\in\aset''$ with $\dset\subseteq\dset''$ and $\dset'\subseteq\dset''$ such that $\dset\cup\dset'\subseteq\dset''$.
			Now let $\bigcup_{i=1}^t(\dset_i\cup\dset_i')$ be an element of $\aset\vee\aset'$.
			Then, $\exists\dset_1'',\ldots,\dset_t''\in\aset''$ with $\dset_1\cup\dset_1'\subseteq\dset_1''$, $\ldots$, $\dset_t\cup\dset_t'\subseteq\dset_t''$.
			Let $i,j\in\{1, \ldots, t\}$ with $i \neq j$.
			If $\dset_i\cup\dset_i'$ and $\dset_j\cup\dset_j'$ are connected by an edge in $G$, then $\dset_i'' = \dset_j''$ because $\aset''$ is a partition of $\rset$.
			Then, since an element of $\aset\vee\aset'$ represents a connected component of $G$, it follows that $\dset_1'' = \ldots = \dset_t''$.
			As a result, for each $\bigcup_{i=1}^t(\dset_i\cup\dset_i')\in\aset\vee\aset'$ $\exists\dset''\in\aset''$ with $\bigcup_{i=1}^t(\dset_i\cup\dset_i')\subseteq\dset''$ and, thus, $\aset\vee\aset'\leq\aset''$.
			Therefore, $\aset\vee\aset'\in\pset_{\rset}$.
		}

		\lemmasupplement{supp_lemma_join_is_smallest_greater}{statements}{
			Let $\crn$ be a chemical reaction network that admits a positive steady state.
			If $\aset,\aset'\in\pset_{\rset}$, then $\aset\vee\aset'$ is the smallest element in $\pset_{\rset}$ which is coarser than or equal to $\aset$ and $\aset'$.
			Therefore, the binary operation $\aset\vee\aset'$ is the \defbf{join} of $\aset, \aset'$.
		}{
			The statement is proved by contradiction:
			Suppose that $\exists\aset''$ with $\aset\leq\aset''$, $\aset'\leq\aset''$ and $\aset''\leq\aset\vee\aset'$.
			Further suppose that $\aset\vee\aset'\neq\aset''$.		
			W.l.o.g., this directly implies that $\exists\bigcup_{i=1}^t(\dset_i\cup\dset_i')\in\aset\vee\aset'$ with $\dset_1,\ldots,\dset_t\in\aset$ and $\dset_1',\ldots,\dset_t'\in\aset'$ and $t\geq2$ such that $\exists\dset''\in\aset''$ with $\dset''\subset\bigcup_{i=1}^t(\dset_i\cup\dset_i')$.
			Then, $\exists i\in\{1,\ldots,t\}$ with $\dset_i\cup\dset_i'\subseteq\dset''$.
			Because $\bigcup_{i=1}^t(\dset_i\cup\dset_i')$ is the union of all nodes of a connected component of $G$, $\exists j\in\{1,\ldots,t\}$ and $i \neq j$ such that $(\dset_i\cup\dset_i')\cap(\dset_j\cup\dset_j')\neq\emptyset$.
			It follows that $\dset_i\cup\dset_i'\cup\dset_j\cup\dset_j'\subseteq\dset''$ since the sets represent equivalence classes of partitions.
			Repeating this argument for all nodes of the corresponding connected component results in $\bigcup_{i=1}^t(\dset_i\cup\dset_i')\subseteq\dset''$ which is a contradiction to $\dset''\subset\bigcup_{i=1}^t(\dset_i\cup\dset_i')$ and, thus, to the assumption.
		}

		\corollarysupplement{supp_corollary_P_R_lattice}{statements}{
			Let $\crn$ be a chemical reaction network that admits a positive steady state.
			The set $\mathcal{P}_{\rset}$ is a lattice.
		}{
			Lemma S\ref{supp_lemma_P_R_is_poset} shows that $\pset_{\rset}$ is a partially ordered set.
			Lemmas S\ref{supp_lemma_meet_is_partition}, S\ref{supp_lemma_meet_in_P_R} and S\ref{supp_lemma_meet_is_greatest_smaller} show that the set defined in Definition S\ref{supp_definition_binary_operation_meet} is the meet of two elements in $\pset_{\rset}$.
			Furthermore, Lemmas S\ref{supp_lemma_join_is_partition}, S\ref{supp_lemma_join_in_P_R} and S\ref{supp_lemma_join_is_smallest_greater} show that the set defined in Definition S\ref{supp_definition_binary_operation_join} is the join of two elements in $\pset_{\rset}$.
			Therefore, $\pset_{\rset}$ is a lattice.
		}

\subsection{$\pset_{\cset}$ is a lattice}
		The set $\pset_{\cset}$ is defined as follows (see the main text for the definition of $\cset/\myrelcsmallx{0}$ and $\mypartc$):

		\definitionsupplement{supp_definition_P_C}{statements}{
			Let $\crn$ be a chemical reaction network that admits a positive steady state.
			Then,
			\begin{equation*}
				\pset_{\cset} := \{\cset/\myrelcsmallx{'} \myspace\vert\myspace \cset/\myrelcsmallx{0}\leq\cset/\myrelcsmallx{'}\myspace\text{and}\myspace\cset/\myrelcsmallx{'}\leq\mypartc\}.
			\end{equation*}\vspace{-0.5cm}
		}
		
		\lemmasupplement{supp_lemma_P_C_is_poset}{statements}{
			Let $\crn$ be a chemical reaction network that admits a positive steady state.
			The set $\mathcal{P}_{\cset}$ is a partially ordered set.
		}{
			The proof works analogously to the proof of Lemma S\ref{supp_lemma_P_R_is_poset}.
		}

		\definitionsupplement{supp_definition_binary_operation_meet_complexes}{statements}{
			Let $\crn$ be a chemical reaction network that admits a positive steady state.
			The binary operation $\mathcal{A} \wedge\mathcal{A}'$ of two elements $\mathcal{A},\mathcal{A}'\in\mathcal{P}_{\mathcal{C}}$ is defined by $\mathcal{A} \wedge\mathcal{A}' := \{\uset \cap \uset' \myspace\vert\myspace \uset\in\mathcal{A}\myspace\text{and}\myspace \uset'\in\mathcal{A}'\myspace\text{and}\myspace \uset\cap\uset' \neq\emptyset\}$.
		}

		\lemmasupplement{supp_lemma_meet_is_partition_complexes}{statements}{
			Let $\crn$ be a chemical reaction network that admits a positive steady state.
			Given two elements $\mathcal{A},\mathcal{A}'\in\mathcal{P}_{\cset}$, the set $\mathcal{A} \wedge\mathcal{A}'$ is a partition of the set of complexes.
		}{
			The proof works analogously to the proof of Lemma S\ref{supp_lemma_meet_is_partition}.
		}

		\lemmasupplement{supp_lemma_meet_in_P_C}{statements}{
			Let $\crn$ be a chemical reaction network that admits a positive steady state.
			Given two elements $\mathcal{A},\mathcal{A}'\in\mathcal{P}_{\cset}$, the set $\mathcal{A} \wedge\mathcal{A}'$ is an element of $\pset_{\cset}$.
		}{
			Lemma S\ref{supp_lemma_meet_is_partition_complexes} shows that $\mathcal{A}\wedge\mathcal{A}'$ is a partition of $\mathcal{C}$.
			The partition $\{\{y\}\myspace\vert\myspace y\in\mathcal{C}\}$ is the finest partition of $\cset$ and is element of $\pset_{\cset}$, i.e., $\{\{y\}\myspace\vert\myspace y\in\mathcal{C}\}\leq\mathcal{A}\wedge\mathcal{A}'$.
			The second part of the proof works analogously to the second part of the proof of Lemma S\ref{supp_lemma_meet_in_P_R}.
		}

 		\lemmasupplement{supp_lemma_meet_is_greatest_smaller_complexes}{statements}{
 			Let $\crn$ be a chemical reaction network that admits a positive steady state.
 			If $\aset,\aset'\in\pset_{\cset}$, then $\aset\wedge\aset'$ is the greatest element in $\pset_{\cset}$ which is finer than or equal to $\aset$ and $\aset'$.
 			Therefore, the binary operation $\aset\wedge\aset'$ is the \defbf{meet} of $\aset, \aset'$.	
 		}{
 			The proof works analogously to the proof of Lemma S\ref{supp_lemma_meet_is_greatest_smaller}.
 		}

		\definitionsupplement{supp_definition_binary_operation_join_complexes}{statements}{
			Let $\crn$ be a chemical reaction network that admits a positive steady state and let $\aset,\aset'\in\pset_{\cset}$.
			Furthermore, let $G := (\vset, \eset)$ be a graph with $\vset = \{\uset\cup\uset'\myvert\uset\in\aset, \myspace\uset'\in\aset', \myand\uset\cap\uset'\neq\emptyset\}$ and $\eset=\{(v,v')\myvert v,v'\in\vset\myand v\cap v'\neq\emptyset\}$.
			Let $G$ be decomposed as $G = G^{(1)} \cup \ldots \cup G^{(s)}$ where $G^{(i)} = (\vset^{(i)},\eset^{(i)})$ with $1\leq i\leq s$ represent the connected components.
			Then, the binary operation $\aset\vee\aset'$ is defined by
			\begin{equation*}
				\aset\vee\aset':=\Biggl\{\bigcup_{v\in\vset^{(i)}} v\myspace\vert\myspace 1\leq i\leq s\Biggr\}.
			\end{equation*}
			\vspace{-0.3cm}
		}

		\lemmasupplement{supp_lemma_join_is_partition_complexes}{statements}{
			Let $\crn$ be a chemical reaction network that admits a positive steady state.
			Given two elements $\aset,\aset'\in\pset_{\rset}$, the set $\aset\vee\aset'$ is a partition of the set of reactions.
		}{
			The proof works analogously to the proof of Lemma S\ref{supp_lemma_join_is_partition}.
		}

		\lemmasupplement{supp_lemma_join_in_P_C}{statements}{
			Let $\crn$ be a chemical reaction network that admits a positive steady state.
			Given two elements $\aset,\aset'\in\pset_{\rset}$, the set $\aset\vee\aset'$ is an element of $\pset_{\cset}$.
		}{
			The proof works analogously to the proof of Lemma S\ref{supp_lemma_join_in_P_R}.
		}

 		\lemmasupplement{supp_lemma_join_is_smallest_greater_complexes}{statements}{
 			Let $\crn$ be a chemical reaction network that admits a positive steady state.
 			If $\aset,\aset'\in\pset_{\cset}$, then $\aset\vee\aset'$ is the smallest element in $\pset_{\cset}$ which is coarser than or equal to $\aset$ and $\aset'$.
 			Therefore, the binary operation $\aset\vee\aset'$ is the \defbf{join} of $\aset, \aset'$.
 		}{
 			The proof works analogously to the proof of Lemma S\ref{supp_lemma_join_is_smallest_greater}.
 		}

		\corollarysupplement{supp_corollary_P_C_lattice}{statements}{
			Let $\crn$ be a chemical reaction network that admits a positive steady state.
			The set $\mathcal{P}_{\cset}$ is a lattice.
		}{
			The proof works analogously to the proof of Lemma S\ref{supp_corollary_P_R_lattice}.
		}
		
	\subsection{$\pset_{\rset}$ is homomorphic to $\pset_{\cset}$}
		In this section, we show that there exists a map $\varphi: \pset_{\rset} \rightarrow \pset_{\cset}$ that preserves the structure of $\pset_{\rset}$.
	
		\definitionsupplement{supp_definition_map_set_of_reactions_to_set_of_complexes}{statements}{
			Let $\crn$ be a chemical reaction network that admits a positive steady state.
			Further, suppose that some partition $\aset\in\pset_{\rset}$ is given.
			The map $\varphi: \pset_{\rset} \rightarrow \pset_{\cset}$ is defined by
			\begin{equation*}
				\varphi(\aset) := \{\uset_{\dset} \myspace\vert\myspace \dset\in\aset\}\cup\{\{y\}\myspace\vert\myspace y\in\overline{\uset}\}.
			\end{equation*}\vspace{-0.75cm}			
		}
	
		First, it is shown that $\varphi(\aset\wedge\aset') = \varphi(\aset)\wedge\varphi(\aset')$ for two elements $\aset,\aset'\in\pset_{\rset}$.
		Let $\dset\in\aset$, $\dset'\in\aset'$ and $\aset,\aset'\in\pset_{\rset}$.
		It is not difficult to see that $\uset_{\dset\cap\dset'} = \{y\myspace\vert\myspace\ri\in\dset\cap\dset'\} = \{y\myspace\vert\myspace\ri\in\dset\}\cap\{y\myspace\vert\myspace\ri\in\dset'\} = \uset_{\dset}\cap\uset_{\dset'}$.
		Since $y\rightarrow\tilde{y}'\in\rset$, $\ri\in\dset$ implies $y\rightarrow\tilde{y}'\in\dset$, it follows that $\dset = \dset_{\uset_{\dset}}$.
		Consequently, $\dset\cap\dset'\neq\emptyset$ if and only if $\uset_{\dset}\cap\uset_{\dset'}\neq\emptyset$. 

		\lemmasupplement{supp_lemma_meet_homomorphism}{statements}{
			Let $\crn$ be a chemical reaction network that admits a positive steady state.
			Then, for two elements $\aset,\aset'\in\pset_{\rset}$, it holds that $\varphi(\aset\wedge\aset') = \varphi(\aset)\wedge\varphi(\aset')$.
		}{
			From the previous arguments, it follows that
			\begin{equation*}
				\equationcolsep
				\begin{array}[h]{rcl}
					\varphi(\aset\wedge\aset')          & = & \{\uset_{\dset\cap\dset'}\myvert\dset\in\aset, \dset'\in\aset', \dset\cap\dset'\neq\emptyset\}\cup\overline{\yset}\\
											            & = & \{\uset_{\dset}\cap\uset_{\dset'}\myvert\dset\in\aset,\dset'\in\aset',\dset\cap\dset'\neq\emptyset\}\cup\overline{\yset}\\
					                                    & = & \{\uset_{\dset}\cap\uset_{\dset'}\myvert\uset_{\dset}\in\varphi(\aset),\uset_{\dset'}\in\varphi(\aset'),\uset_{\dset}\cap\uset_{\dset'}\neq\emptyset\}\cup\overline{\yset}\\
					\varphi(\aset)\wedge\varphi(\aset') & = & \{\uset\cap\uset'\myvert\uset\in\varphi(\aset),\uset'\in\varphi(\aset'),\uset\cap\uset'\neq\emptyset\},
				\end{array}
			\end{equation*}
			which proves the lemma.
		}

		Second, it is shown that $\varphi(\aset\vee\aset') = \varphi(\aset)\vee\varphi(\aset')$ for two elements $\aset,\aset'\in\pset_{\rset}$.
		It is also not difficult to see that $\uset_{\dset\cup\dset'} = \{y\myspace\vert\myspace\ri\in\dset\cup\dset'\} = \{y\myspace\vert\myspace\ri\in\dset\}\cup\{y\myspace\vert\myspace\ri\in\dset'\} = \uset_{\dset}\cup\uset_{\dset'}$.
		
		\lemmasupplement{supp_lemma_join_homomorphism}{statements}{
			Let $\crn$ be a chemical reaction network that admits a positive steady state.
			Then, for two elements $\aset,\aset'\in\pset_{\rset}$, it holds that $\varphi(\aset\vee\aset') = \varphi(\aset)\vee\varphi(\aset')$.
		}{
			Let $G_{\rset} := (\vset_{\rset}, \eset_{\rset})$ be the graph analogous to the graph in Definition S\ref{supp_definition_binary_operation_join}, consisting of $s_{\rset}$ connected components.
			Further, let $G_{\cset} := (\vset_{\cset}, \eset_{\cset})$ be the graph analogous to the graph in Definition S\ref{supp_definition_binary_operation_join_complexes}, consisting of $s_{\cset}$ connected components.
			Then,
			\begin{equation*}
				\equationcolsep
				\begin{array}[h]{rcl}
					\varphi(\aset\vee\aset')          & = & \{\uset_{\bigcup_{v\in\vset^{(i)}_{\rset}}v}\myvert 1\leq i\leq s_{\rset}\}\cup\overline{\yset}\\
					                                  & = & \{\bigcup_{v\in\vset^{(i)}_{\rset}}\uset_v\myvert 1\leq i\leq s_{\rset}\}\cup\overline{\yset}\\
					                                  & = & \{\bigcup_{\dset\cup\dset'\in\vset^{(i)}_{\rset}}\uset_{\dset\cup\dset'}\myvert 1\leq i\leq s_{\rset}\}\cup\overline{\yset}\\					                                  
					                                  & = & \{\bigcup_{\dset\cup\dset'\in\vset^{(i)}_{\rset}} \uset_{\dset}\cup\uset_{\dset'}\myvert 1\leq i\leq s_{\rset}\}\cup\overline{\yset}\\
					                                  & = & \{\bigcup_{\uset_{\dset}\cup\uset_{\dset'}\in\vset^{(i)}_{\cset}} \uset_{\dset}\cup\uset_{\dset'}\myvert 1\leq i\leq s_{\cset}\}\\
					                                  & = & \{\bigcup_{\uset\cup\uset'\in\vset^{(i)}_{\cset}} \uset\cup\uset'\myvert 1\leq i\leq s_{\cset}\}\\
					\varphi(\aset)\vee\varphi(\aset') & = & \{\bigcup_{v\in\vset^{(i)}_{\cset}} v\myvert 1\leq i\leq s_{\cset}\},
				\end{array}
			\end{equation*}
			which proves the lemma.		
		}		

		\corollarysupplement{supp_phi_homomorphism}{statements}{
			Let $\crn$ be a chemical reaction network that admits a positive steady state.
			The map $\varphi: \pset_{\rset} \rightarrow \pset_{\cset}$ is a homomorphism.			
		}{
			Lemmas S\ref{supp_lemma_meet_homomorphism} and S\ref{supp_lemma_join_homomorphism} show that $\varphi(\aset\wedge\aset') = \varphi(\aset)\wedge\varphi(\aset')$ and $\varphi(\aset\vee\aset') = \varphi(\aset)\vee\varphi(\aset')$.
			Therefore, $\varphi$ preserves the structure of $\pset_{\rset}$.
		}

		Analogous to map $\varphi$, there also exists a map $\mu:\pset_{\cset}\rightarrow\pset_{\rset}$ which converts partitions of the set of complexes to partitions of the set of reactions.

		\definition{definition_mapback}{statements}{
			Let $\crn$ be a chemical reaction network that admits a positive steady state.
			Furthermore, suppose a partition $\aset\in\pset_{\cset}$.
			The map $\mu: \pset_{\cset} \rightarrow \pset_{\rset}$ is defined by
			\begin{equation*}
				\mu(\aset) := \{\dset_{\uset} \myspace\vert\myspace \uset\in\aset\}\setminus\emptyset.
			\end{equation*}\vspace{-0.75cm}
		}

		While the map $\varphi$ preserves the structure of $\pset_{\rset}$, this is not true for $\mu$.
		In the following we show that $\mu$ is a surjective function.

		\lemmasupplement{supp_lemma_meet_mu}{statements}{
			Let $\crn$ be a chemical reaction network that admits a positive steady state.
			Then, for two elements $\aset,\aset'\in\pset_{\cset}$, it holds that $\mu(\aset\wedge\aset') = \mu(\aset)\wedge\mu(\aset')$.
		}{
			There are three different cases for $\dset_{\uset\cap\uset'}$ and $\dset_{\uset}\cap\dset_{\uset'}$, respectively, where $\uset\in\aset,\uset'\in\aset'$:
			\begin{itemize}
              \item[(1)] $\uset\cap\uset'\subseteq\overline{\uset}\Rightarrow\dset_{\uset\cap\uset'}=\dset_{\uset}\cap\dset_{\uset'}=\emptyset$.
              \item[(2)] $(\uset\cap\uset')\cap\overline{\uset}\neq\emptyset$ (and not case ($1$)) $\Rightarrow\dset_{\uset\cap\uset'}=\dset_{\uset}\cap\dset_{\uset'}\neq\emptyset$.
              \item[(3)] $(\uset\cap\uset')\cap\overline{\uset}=\emptyset\Rightarrow\dset_{\uset\cap\uset'}=\dset_{\uset}\cap\dset_{\uset'}\neq\emptyset$.
            \end{itemize}
            That is, for cases ($2$) and ($3$), $\dset_{\uset}\in\mu(\aset)$ and $\dset_{\uset'}\in\mu(\aset')$.
  			Then,
  			\begin{equation*}
  				\equationcolsep
 				\begin{array}[h]{rcl}
 					\mu(\aset\wedge\aset')      & = & \{\dset_{\uset\cap\uset'}\myvert\uset\in\aset, \uset'\in\aset', \uset\cap\uset'\neq\emptyset\}\setminus\emptyset\\
 										        & = & \{\dset_{\uset}\cap\dset_{\uset'}\myvert\uset\in\aset,\uset'\in\aset',\uset\cap\uset'\neq\emptyset\}\setminus\emptyset\\
 					\mu(\aset)\wedge\mu(\aset') & = & \{\dset\cap\dset'\myvert\dset\in\mu(\aset),\dset'\in\mu(\aset'),\dset\cap\dset'\neq\emptyset\},
  				\end{array}
  			\end{equation*}
  			which proves the lemma.		
		}
		
		For the remaining proof we need some additional definitions.
		Let $\overline{\vset}_{\cset}$ be the set of nodes in $G_{\cset}$ with $\overline{\vset}_{\cset} := \{v\in\vset_{\cset}\myvert v\subseteq\overline{\uset}\}$.
		Analogously, let $\overline{\eset}_{\cset}$ be the set of edges in $G_{\cset}$ with $\overline{\eset}_{\cset} := \{(u,v)\in\eset_{\cset}\myvert u\cap v\subseteq\overline{\uset}\}$.		
		Finally, let $\tilde{G}_{\cset} := (\vset_{\cset}\setminus\overline{\vset}_{\cset}, \eset_{\cset}\setminus\overline{\eset}_{\cset}) = (\tilde{\vset}_{\cset}, \tilde{\eset}_{\cset}) = \{\tilde{G}_{\cset}^{(1)},\ldots,\tilde{G}_{\cset}^{(t)}\}$ with $\tilde{G}_{\cset}^{(i)} := (\tilde{\vset}_{\cset}^{(i)},\tilde{\eset}_{\cset}^{(i)})$ being the connected components of $\tilde{G}_{\cset}$.
		It is not difficult to see that $\tilde{G}_{\cset}$ represents a partition of the set of substrate complexes.
		Since each substrate complex corresponds to a set of reactions which are elements of the same equivalence class of a partition in $\pset_{\rset}$, such a partition of the set of substrate complexes is equivalent to a partition of the set of reactions.
		Then, it directly follows that $\rset/\myrelrsmallx{''}\leq\rset/\myrelrsmallx{'}$, where $\rset/\myrelrsmallx{'} = \mu(\{\bigcup_{v\in\vset_{\cset}^{(i)}}\myvert 1\leq i\leq s_{\cset}\})$ and $\rset/\myrelrsmallx{''} = \mu(\{\bigcup_{v\in\tilde{\vset}_{\cset}^{(i)}}\myvert 1\leq i\leq t\})$.

		\lemmasupplement{no_name_yet}{statements}{
			Let $\crn$ be a chemical reaction network that admits a positive steady state.
			Further, let two elements $\aset,\aset'\in\pset_{\cset}$ be given such that, based on $\aset\vee\aset'$, the graph $\tilde{G}_{\cset}$ is defined and, based on $\mu(\aset)\vee\mu(\aset')$, the graph $G_{\rset}$ is defined.
			If $\rset/\myrelrsmallx{''} = \mu(\{\bigcup_{v\in\tilde{\vset}_{\cset}^{(i)}}\myvert 1\leq i\leq t\})$ and $\rset/\myrelrsmallx{'''} = \{\bigcup_{v\in\vset_{\rset}^{(i)}}\myvert 1\leq i\leq s_{\rset}\}$, then $\rset/\myrelrsmallx{''} = \rset/\myrelrsmallx{'''}$.
		}{
			The proof consists of two parts:
			\begin{itemize}
				\item[(1)] $\Rightarrow$
					\begin{itemize}
						\item[(1.i)]   $\uset\cup\uset'\in\tilde{\vset}_{\cset}^{(i)}\Rightarrow\uset\cap\uset'\neq\emptyset\not\subseteq\overline{\uset}\Rightarrow\dset_{\uset}\cap\dset_{\uset'}\neq\emptyset\Rightarrow\exists j\text{\ with\ }\dset_{\uset}\cup\dset_{\uset'}\in\vset_{\rset}^{(j)}$ since $\uset\in\aset, \uset'\in\aset'$ implies $\dset_{\uset}\in\mu(\aset), \dset_{\uset'}\in\mu(\aset')$.
						\item[(1.ii)]  $((\uset\cup\uset'),(\tilde{\uset}\cup\tilde{\uset}'))\in\tilde{\eset}_{\cset}^{(i)}\Rightarrow(\uset\cup\uset')\cap(\tilde{\uset}\cup\tilde{\uset}')\neq\emptyset\not\subseteq\overline{\uset}\Rightarrow(\dset_{\uset}\cup\dset_{\uset'})\cap(\dset_{\tilde{\uset}}\cup\dset_{\tilde{\uset}'})\neq\emptyset\Rightarrow\exists j\text{\ with\ }((\dset_{\uset}\cup\dset_{\uset'}),(\dset_{\tilde{\uset}}\cup\dset_{\tilde{\uset}'}))\in\eset_{\rset}^{(j)}\Rightarrow\dset_{\tilde{\uset}}\cup\dset_{\tilde{\uset}'}\in\vset_{\rset}^{(j)}$.
						\item[(1.iii)] Finally, from (1.i) and (1.ii) follows $\{\dset_{\uset\cup\uset'}\myvert\uset\cup\uset'\in\tilde{\vset}_{\cset}^{(i)}\}\subseteq\vset_{\rset}^{(j)}$.
                    \end{itemize}
            	\item[(2)] $\Leftarrow$
					\begin{itemize}
             			\item[(2.i)]  $\dset\cup\dset'\in\vset_{\rset}^{(j)}\Rightarrow\dset\cap\dset'\neq\emptyset\Rightarrow\uset_{\dset}\cap\uset_{\dset'}\neq\emptyset\Rightarrow\exists k\text{\ with\ }\uset_{\dset}\cup\uset_{\dset'}\in\tilde{\vset}_{\cset}^{(k)}$ since $\dset\in\mu(\aset), \dset'\in\mu(\aset')$ implies $\uset_{\dset}\in\aset, \uset_{\dset'}\in\aset'$.
						\item[(2.ii)] $((\dset\cup\dset'),(\tilde{\dset}\cup\tilde{\dset}'))\in\eset_{\rset}^{(j)}\Rightarrow(\dset\cup\dset')\cap(\tilde{\dset}\cup\tilde{\dset}')\neq\emptyset\Rightarrow(\uset_{\dset}\cup\uset_{\dset'})\cap(\uset_{\tilde{\dset}}\cup\uset_{\tilde{\dset}'})\neq\emptyset\not\subseteq\overline{\uset}\Rightarrow\exists k\text{\ with\ }((\uset_{\dset}\cup\uset_{\dset'}),(\uset_{\tilde{\dset}}\cup\uset_{\tilde{\dset}'}))\in\tilde{\eset}_{\cset}^{(k)}\Rightarrow\uset_{\tilde{\dset}}\cup\uset_{\tilde{\dset}'}\in\tilde{\vset}_{\cset}^{(k)}$.
						\item[(2.iii)] Finally, from (2.i) and (2.ii) follows $\vset_{\rset}^{(j)}\subseteq\{\dset_{\uset\cup\uset'}\myvert\uset\cup\uset'\in\tilde{\vset}_{\cset}^{(k)}\}$.
            		\end{itemize}
            \end{itemize}
            Then, since $\tilde{G}_{\cset}$ is equivalent to a partition of the set of reactions, from (1.iii) and (2.iii) it follows that $\tilde{\vset}_{\cset}^{(i)} = \tilde{\vset}_{\cset}^{(k)}$.
            Therefore, $\forall i\in\{1,\ldots, t\}$ $\exists j\in\{1,\ldots, s_{\rset}\}$ (and vice versa) with $\{\dset_{\uset\cup\uset'}\myvert\uset\cup\uset'\in\tilde{\vset}_{\cset}^{(i)}\}=\vset_{\rset}^{(j)}$.
		}
		
		\corollarysupplement{supp_lemma_join_mu}{statements}{
			Let $\crn$ be a chemical reaction network that admits a positive steady state.
			Then, for two elements $\aset,\aset'\in\pset_{\cset}$, it holds that $\mu(\aset)\vee\mu(\aset')\leq\mu(\aset\vee\aset')$.
		}{
			Lemma S\ref{no_name_yet} shows that $\mu(\aset)\vee\mu(\aset') = \mu(\{\bigcup_{v\in\tilde{\vset}_{\cset}^{(i)}}\myvert 1\leq i\leq t\})$.
			As a result, $\mu(\aset)\vee\mu(\aset') = \mu(\{\bigcup_{v\in\tilde{\vset}_{\cset}^{(i)}}\myvert 1\leq i\leq t\}) \leq \mu(\{\bigcup_{v\in\vset_{\cset}^{(i)}}\myvert 1\leq i\leq s_{\cset}\}) = \mu(\aset\vee\aset')$.
		}		

\newpage
\bibliographystyle{apalike}
\bibliography{my_bibliography}
\end{document}